\documentclass{article}

\usepackage{spconf,graphicx}
\usepackage{cite} 
\usepackage{subfigure} 
\usepackage{color}
\usepackage[dvipsnames,  x11names, table]{xcolor}
\usepackage{amsmath,amssymb,amsfonts}
\usepackage{tabularx}
\usepackage{color}
\usepackage{cite}
\usepackage{pstool}
\usepackage{psfrag}
\usepackage{lettrine}
\usepackage{float}
\usepackage[export]{adjustbox}
\usepackage[caption = false]{subfig}
\usepackage{booktabs}
\usepackage[dvips]{epsfig}
\usepackage{multirow}
\usepackage{kotex}
\usepackage[ruled,vlined,linesnumbered,noresetcount]{algorithm2e}

\DeclareMathOperator*{\argmax}{argmax}
\DeclareMathOperator*{\argmin}{argmin}

\title{Experimental Demonstration of Location-aware Beam Alignment}

\name{Junyeol Hong \qquad Hyeonjin Chung \qquad Sunwoo Kim}
\address{Department of Electronics and Computer Engineering, Hanyang University, Seoul, Korea \\
Email: \{ftsh610, hyeonjingo, remero\}@hanyang.ac.kr}

\begin{document}

\maketitle

\begin{abstract}
The main focus of beam alignment is to find the optimal beam which yields the largest received signal strength (RSS) with faster speed.
In this paper, we demonstrate an efficient beam alignment scheme with our testbed. The algorithm we experiment uses the location information for the computation efficient beam alignment.
The testbed transmits and receives the 13.8 GHz signal and steers a beam on both transmitter and receiver with various radio frequency (RF) components. The location information is estimated with the indoor positioning module. 
The experiment shows that the location-aware algorithm significantly reduces the time consumption for beam alignment than the exhaustive search.


\end{abstract}


\section{Introduction}
Millimeter-wave (mmWave) has become one of the key enabler for achieving higher data rate~\cite{6515173}. However, the high path loss and the severe scattering lower the received signal strength (RSS), which makes mmWave communication challenging~\cite{1491267}. For this reason, beamforming is widely employed for mmWave systems with an array antenna and a network of phase shifters~\cite{6736750}. Beamforming creates a highly directional radiation pattern, thus the beam between a base station (BS) and a user equipment (UE) must be aligned. This has invoked a new study for beam management, the procedure for establishing and maintaining the beam alignment.

The focus of the beam management is to make beam aligned with faster speed, yet accurate. The method that is used by current standard forms a beam at every direction, then find the beam with the largest RSS~\cite{7945855,8258595}. It is widely known as exhaustive search, whose robustness is guaranteed since it searches all possible directions. However, it has crucial disadvantage in large time consumption. 
To reduce the time consumption for beam alignment, the approach that utilizes the location information has been proposed~\cite{7536855,7786130,Orikumhi2019LocationawareBA}. Assuming the noisy location is given, it creates a light codebook whose size is proportional with the uncertainty of the location. Location-aware method can reduce the time consumption dramatically as the positioning accuracy increases.     

The realistic implementations of beam management are done in~\cite{7794570,7343415,8464878,8690621,8385478}. Aforementioned works in~\cite{7794570,7343415} focus on the validation of the next generation mobile communication on the realistic environment, and thus they employ the exhaustive search which conforms with the standard. Indoor experiments on 60 GHz are done in~\cite{8464878,8690621}, still they use the exhaustive search for beam alignment. Realistic experiment of other beam alignment algorithm such as hierarchical search~\cite{Hier} is conducted in~\cite{8385478}. However, the additional information such as location is not used for beam alignment.  

In this paper, we implement the location-aware beam alignment algorithm using the testbed, where the concept of the algorithm is based on~\cite{Orikumhi2019LocationawareBA}. 
The algorithm is demonstrated on the realistic 13.8 GHz beam tracking testbed, and the experiment shows that the proposed algorithm reduces the time consumption for beam alignment compared to the exhaustive search.


\section{Location-aware Beam Alignment}\label{2}
We use two antenna arrays on each of transmit side and receive side which forms multiple-input-multiple-output (MIMO) channel. 
We assume the array has $M$ elements and its spacing between adjacent antennas equals to a half-wavelength of the carrier frequency.
The array manifold vector is defined as $\mathbf{a}(\theta)$, where $\theta$ may denote the angle of arrival (AoA) or the angle of departure (AoD) depending on the situation. 

Since we always secure the dominant line-of-sight (LoS) path in the experiment, the MIMO channel of the $k$-th time slot $\mathbf{H}_{k}$ is given as
\begin{equation}
    \mathbf{H}_{k} \approx \alpha_{k} \mathbf{a}(\theta_{k}) \mathbf{a}(\phi_{k})^{H} \in \mathbb{C}^{M \times M},
\end{equation}
where $\alpha_{k}$, $\theta_{k}$, $\phi_{k}$ respectively represent the channel gain, the AoA, and AoD of the $k$-th time slot. The auxiliary paths are neglected. 

With the application of the beamforming on both sides, a vector of the received signal $\mathbf{y}_{k}$ can be given as
\begin{equation}
    \mathbf{y}_{k}=\mathbf{w}^{H}_{k} \mathbf{H}_{k} \mathbf{f}_{k} \mathbf{s}_{k} + \mathbf{n}_{k} \in \mathbb{C}^{1 \times T},
\end{equation}
where $\mathbf{w}_{k}$ and $\mathbf{f}_{k}$ are the beamforming vector of receiver and transmitter. $\mathbf{s}_{k}$, $\mathbf{n}_{k}$, and $T$ denote a vector of the transmitted signal, the additive white Gaussian noise, and the number of samples used for the RSS measurement. 

The purpose of the demonstration is to find $\mathbf{w}_{k}$ and $\mathbf{f}_{k}$ that maximize $\mathbb{E} \left[ \mathbf{y}_{k} \mathbf{y}^{H}_{k} \right]$ which denotes the RSS. We assume that $\mathbb{E} \left[ \mathbf{s}_{k} \mathbf{s}^{H}_{k} \right]$ remain constant over time and $\mathbb{E} \left[ \mathbf{y}_{k} \mathbf{y}^{H}_{k} \right] \approx \mathbf{y}_{k} \mathbf{y}^{H}_{k}/T$. Then, the equivalent equation for finding adequate beamforming vectors can be summed up to
\begin{equation}
    	\left\{ \hat{\mathbf{w}}_{k},\hat{\mathbf{f}}_{k} \right\}=\argmax_{\mathbf{w}_{k} \in \mathcal{W}_{k}, \mathbf{f}_{k} \in \mathcal{F}_{k}}{\left| \mathbf{w}^{H}_{k} \mathbf{H}_{k} \mathbf{f}_{k} \right|},
\end{equation}
where $\hat{\mathbf{w}}_{k}$ and $\hat{\mathbf{f}}_{k}$ denote the optimal beamforming vectors on both side. $\mathcal{W}_{k}$ and $\mathcal{F}_{k}$ are codebooks which gather the candidates of the beamforming vectors. Without any location information, $\mathcal{W}_{k}=\mathcal{F}_{k}=\left\{\mathbf{a}(\vartheta_{1}),\ldots,\mathbf{a}(\vartheta_{N}) \right\}$, where $\vartheta_n$ denotes the $n$-th pre-defined angle, $\{ 1,\ldots,N \}$ are defined as beam indices, and $N$ denotes the size of the original codebook. 
\begin{figure}[t]
    \begin{center}
        \psfrag{a}[Bc][bc][0.7][0]{Tx}
        \psfrag{b}[Bc][bc][0.6][0]{\shortstack[c]{Rx at the\\$(k{\text{-}}1)$-th time slot}}
        \psfrag{x1}[Bc][bc][0.8][0]{6 m}
        \psfrag{x2}[Bc][bc][0.8][0]{6 m}
        \psfrag{x3}[Bc][bc][0.8][0]{9 m}
        \psfrag{r}[Bc][bc][0.9][0]{Room}
        \psfrag{pa}[Bc][bc][0.9][0]{Positioning Area}
        \psfrag{p}[Bl][bc][0.8][0]{: Indoor positioning module (Pozyx)}
        \psfrag{c2}[Bc][bc][0.8][0]{{$\bar{\phi}_{k}$}}
        \psfrag{c3}[Bc][bc][0.7][0]{\shortstack[c]{\color{red}{Maximum steerable range:}\\ \color{red}{$-30^{\circ} \sim +30^{\circ}$}}}
        \psfrag{d2}[Bc][bc][0.8][0]{{$\bar{\theta}_{k}$}}
        \psfrag{o}[Bc][bc][0.6][0]{\shortstack[c]{Rx at the\\$k$-th time slot}}
        \psfrag{e}[Bc][bc][0.65][0]{$\mathbf{l}^{\textrm{Tx}}_{k}= \left[{l}^{\textrm{Tx}}_{x,k}, {l}^{\textrm{Tx}}_{y,k}\right]$}
        \psfrag{f}[Bc][bc][0.65][0]{$\mathbf{l}^{\textrm{Rx}}_{k{\text{-}}1}= \left[{l}^{\textrm{Rx}}_{x,k{\text{-}}1}, {l}^{\textrm{Rx}}_{y,k{\text{-}}1}\right]$}
        \psfrag{h}[Bc][bc][0.65][0]{$\mathbf{l}^{\textrm{Rx}}_{k}= \left[{l}^{\textrm{Rx}}_{x,k}, {l}^{\textrm{Rx}}_{y,k}\right]$}
        \includegraphics[width=1\columnwidth]{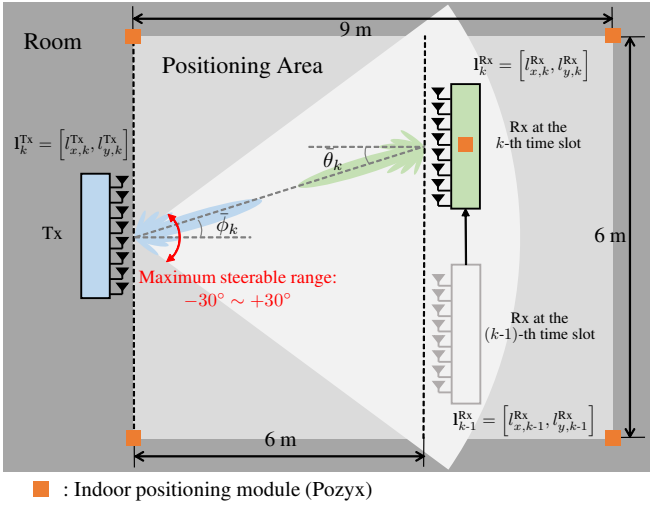}
        \caption{A geometric representation of the experiment environment.}
        \label{environments}
    \end{center}
\end{figure}
However, candidates for the beam alignment can be narrowed with the location information since it provides the AoA and the AoD. The location of the transmitter and the receiver at the $k$-th time slot, $\mathbf{l}^{\textrm{Tx}}_{k}$ and $\mathbf{l}^\textrm{Rx}_{k}$ are
\begin{equation}
    \mathbf{l}^{\textrm{Tx}}_{k} = \left[{l}^{\textrm{Tx}}_{x,k}, {l}^{\textrm{Tx}}_{y,k}\right],\;
    \mathbf{l}^{\textrm{Rx}}_{k} = \left[{l}^\textrm{Rx}_{x,k}, {l}^\textrm{Rx}_{y,k}\right],
\end{equation}
where ${l}^{\textrm{Tx}}_{x,k}$ and ${l}^{\textrm{Tx}}_{y,k}$ denote coordinates of the transmitter on the x-axis and the y-axis, while 
${l}^{\textrm{Rx}}_{x,k}$ and ${l}^{\textrm{Rx}}_{y,k}$ denote coordinates of the receiver on the x-axis and the y-axis.
In the experiment, the transmitter and the receiver face each other as in Fig.~\ref{environments}, such that the orientation of both side are equal. In this case, desirable beam steering directions of both sides, $\bar{\phi}_{k}$ and $\bar{\theta}_{k}$ can be represented as 
\begin{equation}\label{best_angle}
    \bar{\phi}_{k} = \bar{\theta}_{k} = \arctan\left(\frac{{{l}^\textrm{Rx}_{y,k}} - {l}^{\textrm{Tx}}_{y,k} }{{{l}^\textrm{Rx}_{x,k}} - {l}^{\textrm{Tx}}_{x,k}}\right).
\end{equation}
Then, we select the best beam indices as follows.
\begin{equation}\label{beamidx}
    P_{k}=\argmin_{ i \in \{ 1,\ldots,N \} }{\left| \vartheta_{i}-\bar{\phi}_{k} \right|}, \;
    Q_{k}=\argmin_{ i \in \{ 1,\ldots,N \} }{\left| \vartheta_{i}-\bar{\theta}_{k} \right|},
\end{equation}
where $P_{k}$ and $Q_{k}$ denote the best indices for the transmitter and the receiver.  

However, we cannot use $\vartheta_{{P_{k}}}$ and $\vartheta_{{Q_{k}}}$ directly as the best beam pair for hardware-induced problem.
Radio frequency (RF) components such as power splitter, phase shifter, and cable may cause the undesired phase unbalance.
Additionally, the radiation pattern of each antenna element varies. These problems cause a distortion of the beam pattern, and their influence on system is hard to estimate without precise measurement.

For this reason, we partially search beams that are adjacent to the $P_{k}$-th beam for the transmitter and the $Q_{k}$-th beam for the receiver.
In our environment, it is reasonable to search only adjacent beams, considering that the hardware-induced factors do not distort the main beam direction more than $5^{\circ}$.
Then, $\mathcal{W}_{k}$ and $\mathcal{F}_{k}$ can be given as   


\begin{equation}\label{reduced}
\begin{split}
    {\mathcal{W}}_{k} &= \{\mathbf{a}(\vartheta_{{Q_{k}}-1}),\mathbf{a}(\vartheta_{{Q_{k}}}),\mathbf{a}(\vartheta_{{Q_{k}}+1})\},\\
    {\mathcal{F}}_{k} &= \{\mathbf{a}(\vartheta_{{P_{k}}-1}),\mathbf{a}(\vartheta_{{P_{k}}}),\mathbf{a}(\vartheta_{{P_{k}}+1})\}.
    \end{split}
\end{equation}


\section{Experimental Setup}

\begin{figure}
    \begin{center}
        \psfrag{A1}[Bc][bc][0.7][0]{Analog Beamformer}
        \psfrag{A2}[Bc][bc][0.7][0]{ADAR-1000 Evaluation Board, ADI.}
        \psfrag{B1}[Bc][bc][0.7][0]{ULA Antenna}
        \psfrag{B2}[Bc][bc][0.45][0]{SAM-1431430695-SF-L1-8C}
        \psfrag{B3}[Bc][bc][0.55][0]{Sage Millimeter, Inc.}
        \psfrag{C}[Bc][bc][0.65][0]{Laptop}
        \psfrag{D1}[Bc][bc][0.65][0]{National Inst.}
        \psfrag{D2}[Bc][bc][0.65][0]{NI USRP-2922}
        \psfrag{E}[Bc][bc][0.8][0]{1 Set of test-bed}
        \psfrag{F1}[Bc][bc][0.7][0]{Frequency Converter}
        \psfrag{F2}[Bc][bc][0.7][0]{VCO}
        \includegraphics[width=1\columnwidth]{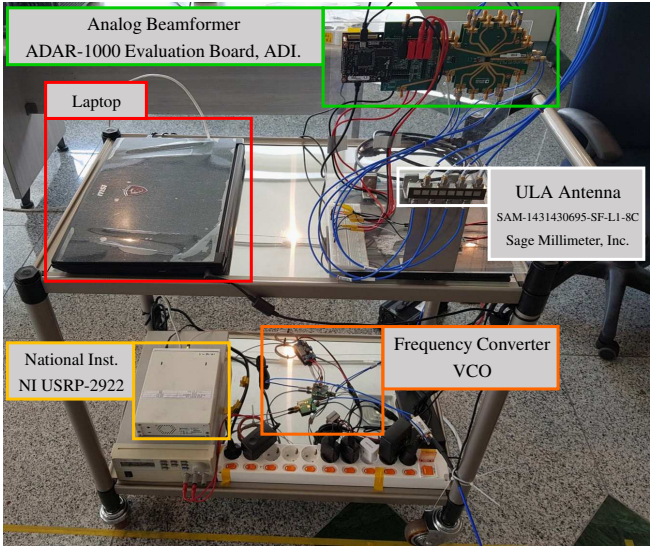}
    \caption{A picture of the testbed.} \label{hardware}
    \end{center}
\end{figure}

The experiments are conducted in a rectangular shaped room whose size is 6 m by 9 m.
Fig.~\ref{environments} shows the geometric representation of the experiment environment. The transmitter and 4 indoor positioning modules are fixed, and one positioning module is attached to the receiver. Fixed positioning modules work as the anchors, while the module on the receiver works as the agent. A picture of the testbed is given in Fig.~\ref{hardware}.

\begin{figure}[t]
    \begin{center}
        \psfrag{a}[Bc][bc][0.7][0]{USRP}
        \psfrag{b}[Bc][bc][0.7][0]{VCO}
        \psfrag{c}[Bc][bc][0.6][0]{Propagation}
        \psfrag{o}[Bc][bc][0.65][0]{Analog beamformer}
        \psfrag{c1}[Bc][bc][0.5][0]{\shortstack[c]{Up\\converter}}
        \psfrag{c2}[Bc][bc][0.5][0]{\shortstack[c]{Down\\converter}}
        \psfrag{d1}[Bc][bc][0.5][0]{2.8 GHz}
        \psfrag{d2}[Bc][bc][0.5][0]{11 GHz}
        \psfrag{u}[Bc][bc][0.5][0]{\shortstack[c]{13.8\\ GHz}}
        \psfrag{t}[Bc][bc][0.8][0]{\textcolor{ForestGreen}{Transmitter}}
        \psfrag{r}[Bc][bc][0.8][0]{\textcolor{Orange}{Receiver}}
        \psfrag{e1}[Bl][bc][0.8][0]{: RF amplifier}
        \psfrag{e2}[Bl][bc][0.8][0]{: Phase shifter}
        \includegraphics[width=1\columnwidth]{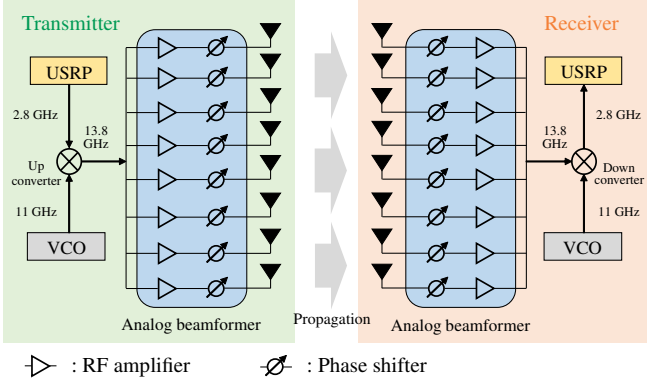}
    \end{center}
    \caption{The architecture of 13.8 GHz beamforming testbed.}
    \label{signal chain}
\end{figure} 

\subsection{Hardware Setup}
The overall structure of the hardware and the flow of the signal is given in Fig.~\ref{signal chain}. The description for each component is given as follows.

\hspace{-\topsep}
\vspace{-\topsep}
\begin{itemize}
    \setlength\itemsep{0em}
    \item{\textit{Universal software radio peripheral (USRP) --} 
    We use 2 NI USRP-2922 at both sides. The USRP at the transmitter generates a signal whose carrier frequency is 2.8 GHz, while the USRP at the receiver analyzes the received signal. 
    }
    \item{\textit{Up/down-converter --} Converters multiply the input signal with 11 GHz sinusoid signal generated by the external voltage controlled oscillator (VCO). The up-converter yields 13.8 GHz signal, and the down-converter yields 2.8 GHz signal.}
    \item{\textit{Analog beamformer --} For the analog beamformer, we use 2 ADAR-1000 evaluation boards produced by Analog Devices, Inc. The analog beamformer consists of a network of the RF amplifiers and the phase shifters.}
    
    \item{\textit{Uniform linear array (ULA) antenna --} ULA antenna is produced by SAGE Millimeter, Inc. It has 8 antenna elements, such that $M=8$, and the spacing between adjacent elements is set to a half-wavelength of 13.8 GHz.}
    
    \item{\textit{Indoor positioning module --} To estimate $\mathbf{l}^{\textrm{Rx}}_{k}$, we use the indoor positioning module, Pozyx. It uses ultra wide band (UWB)-based two-way ranging (TWR) to estimate the position. 5 modules are used to estimate the position of the receiver.
}
\end{itemize}
\vspace{-\topsep}
\hspace{-\topsep}

\subsection{Software Setup}
\begin{figure}[t]
    \begin{center}
    \psfrag{a}[Bc][bc][0.8][0]{Python application}
    \psfrag{b}[Bc][bc][0.8][0]{C\# application}
    \psfrag{c}[Bc][bc][0.8][0]{LabVIEW}
    \psfrag{1}[Bc][bc][0.8][0]{Start}
    \psfrag{2}[Bc][bc][0.8][0]{$k=0$}
    \psfrag{3}[Bc][bc][0.8][0]{\shortstack[c]{Estimate $\mathbf{l}^{\textrm{Rx}}_{k}$ using \\positioning modules}}
    \psfrag{4}[Bc][bc][0.8][0]{\shortstack[c]{Calculate $\bar{\phi}_{k}$, $\bar{\theta}_{k}$\\ and derive $P_{k}$, $Q_{k}$}}
    \psfrag{5}[Bc][bc][0.8][0]{\shortstack[c]{Set $\mathcal{W}_{k}$, $\mathcal{F}_{k}$\\ based on $P_{k}$, $Q_{k}$}}
    \psfrag{6}[Bc][bc][0.8][0]{\shortstack[c]{Sequentially steer \\ beam using $\mathcal{W}_{k}$, $\mathcal{F}_{k}$}}
    \psfrag{7}[Bc][bc][0.8][0]{\shortstack[c]{Measure RSS of \\ every beam pair}}
    \psfrag{8}[Bc][bc][0.8][0]{\shortstack[c]{Find $\{ \hat{\mathbf{w}}_{k},\hat{\mathbf{f}}_{k} \}$}}
    \psfrag{9}[Bc][bc][0.8][0]{\shortstack[c]{$k \leftarrow k+1$}}
    \includegraphics[width=1\columnwidth]{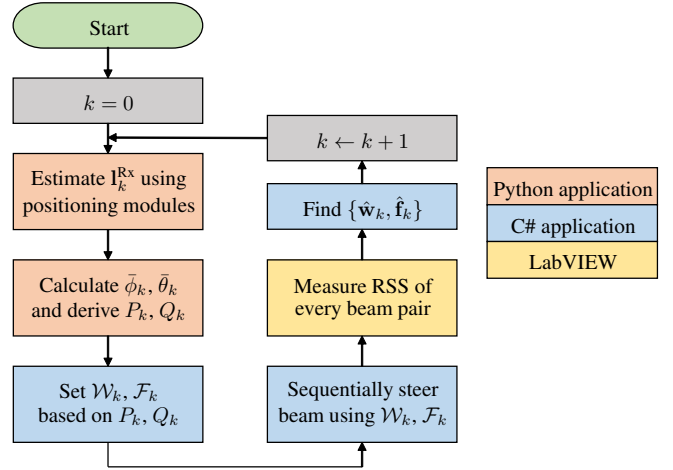}
    \caption{A flowchart of the location-aware beam alignment.} \label{angles}
    \end{center}
\end{figure}

There are mainly 3 software. NI LabVIEW and C\# application run on both sides for controlling USRP and the analog beamformer. Python application runs on the receiver for estimating the position of receiver. The flowchart of the location-aware beam alignment is depicted in Fig.~\ref{angles}, and the role of each software is given as follows. 



\vspace{-\topsep}
\begin{itemize}
    \setlength\itemsep{0em}
    \item{\textit{Python application --} Python application receives the raw range data from positioning modules and estimates $\mathbf{l}^{\textrm{Rx}}_{k}$ using trilateration~\cite{1458275}. Then, Python application reports $\mathbf{l}^{\textrm{Rx}}_{k}$ to C\# application.
    }
    
    \item{\textit{LabVIEW --}  
    LabVIEW on the transmitter generates random bits and modulates them using quadrature phase shift keying (QPSK). Then, the signal is multiplexed by the orthogonal frequency division multiplexing (OFDM) and is handed over to USRP.
    LabVIEW on the receiver demodulates the signal and measures the RSS that corresponds to each beam pair. The RSS information is reported to C\# application. }
    
    \item{\textit{C\# application --} The main role of C\# application is to control analog beamformers to steer beam. It calculates $\bar{\phi}_{k}$ and $\bar{\theta}_{k}$ by (\ref{best_angle}), then it derives $P_{k}$ and $Q_{k}$ by (\ref{beamidx}).
    Both sides share the best beam pair and use it until the next searching process.}
    
\end{itemize}
\vspace{-\topsep}


\section{Experiment Results}
\begin{figure}[t]
    \begin{center}
    \includegraphics[width=1\columnwidth]{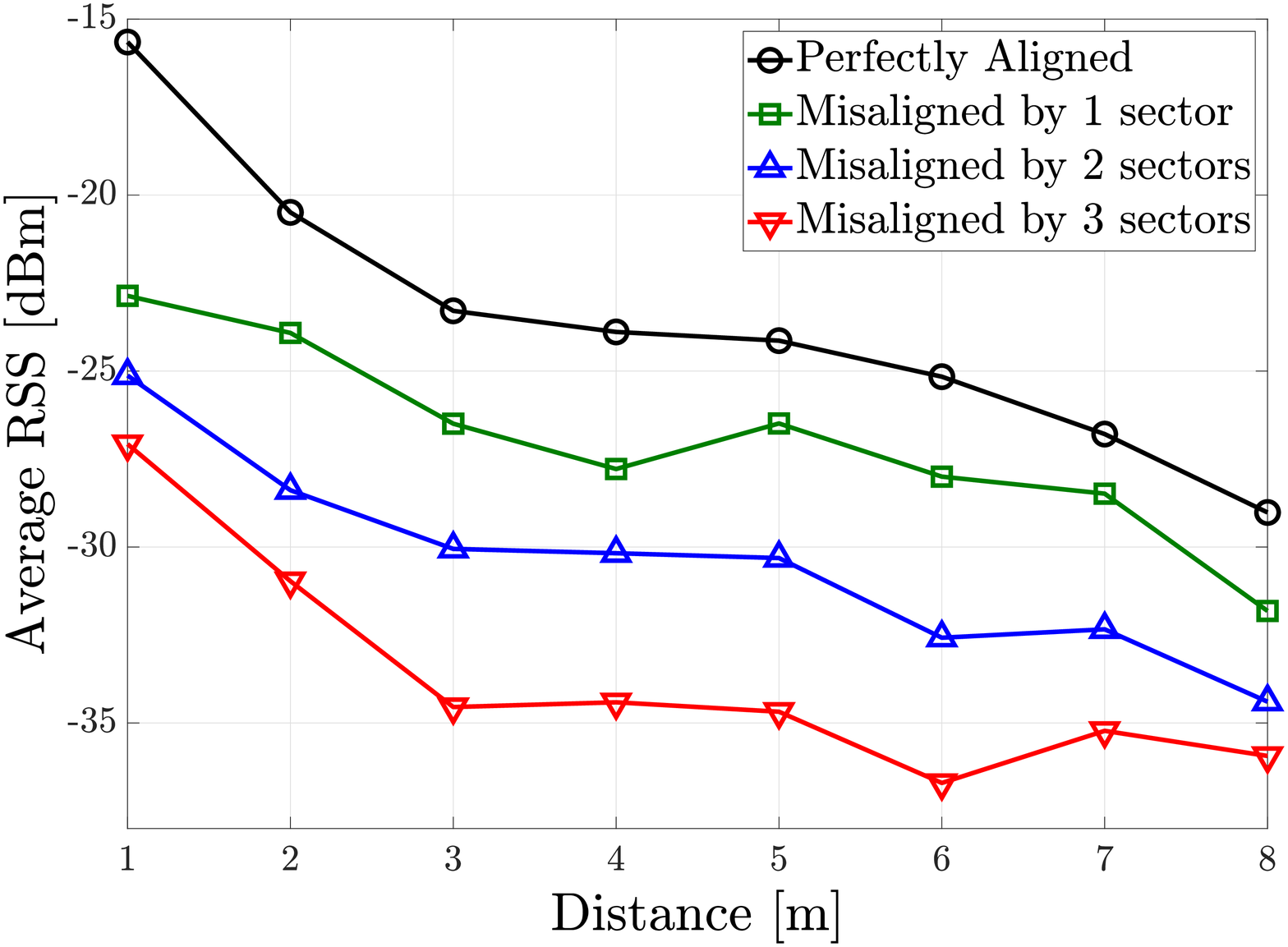}
    \caption{RSS versus distance. The RSS is measured at 1 m interval from 1 m to 8 m.} \label{RSS}
    \end{center}
\end{figure}

We present two experiment results. In the first experiment, $N=7$ and $\vartheta_{n}=\{ -30+10(n-1) \}^{\circ}$. The range of $\vartheta_{n}$ is $[-30^\circ,+30^\circ]$ since the maximum steerable range is limited due to the size of the room as depicted in Fig.~\ref{environments}. 
The first experiment compares the RSS when the beams are aligned and the RSS when they are not. Also, the RSS is measured for the different degrees of misalignment, where the degree of misalignment is represented as the difference of beam index. For example, if the optimal beam pair uses the $a$-th beam for the transmitter and the the $b$-th beam for the receiver, we call the beam is misaligned by $(i+j)$ sectors when the current beam pair uses the $(a \pm i)$-th beam and the the $(b \pm j)$-th beam. In this case, there can be multiple alternatives that represent the equivalent degree of misalignment. Thus, we measure the RSS of all alternatives and average them.  
Fig.~\ref{RSS} shows that the RSS is inversely proportional to the degree of misalignment, where the RSS reaches its peak when the beams are perfectly aligned.   

\begin{figure}[t]
    \begin{center}
    \includegraphics[width=1\columnwidth]{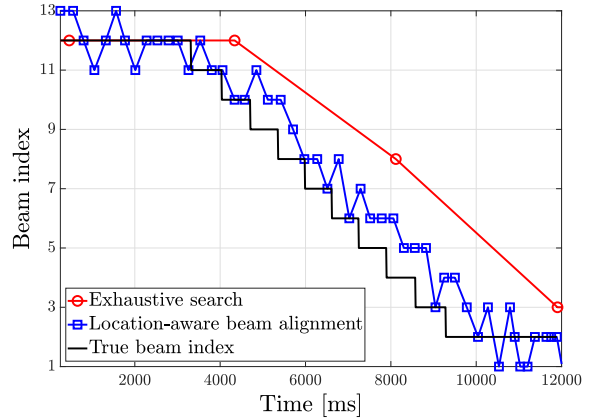}
    \caption{Change in the beam index throughout time.} \label{Beam}
    \end{center}
\end{figure}

In the second experiment, $N=13$ and $\vartheta_{n}=\{ -30+5(n-1) \}^{\circ}$. 
The second experiment shows how the beam index changes when the receiver is mobile. We compare the location-aware beam alignment with the exhaustive search. For the first 3 seconds, the receiver remains motionless. Then, the receiver moves at 3 km/h along with a straight line which is 6 m apart from the transmitter as in Fig.~\ref{environments}. The black line in Fig.~\ref{Beam} denotes how the beam index changes with the real movement. Thus, if one algorithm behaves similarly as the black line in Fig.~\ref{Beam}, we can tell that it successfully tracks the mobility of the receiver.     
Fig.~\ref{Beam} shows that the location-aware beam alignment updates the beam more frequently than the exhaustive search, and thus shows a similar result with the true beam index.
On average, the location-aware beam alignment consumes 0.28 seconds to determine the optimal beams, while the exhaustive search consumes 3.7 seconds.  



\section{Discussion}
The scenario in~\cite{Orikumhi2019LocationawareBA} is based on the outdoor environment with massive MIMO. Both BS and UE use up to 64 antennas, and they produce a beam whose half power beamwidth (HPBW) is $1.6^{\circ}$. 
The scenario in~\cite{Orikumhi2019LocationawareBA} assumes that the distance between the BS and the UE is 100 m.
Given the distance and the HPBW, the region that is covered by the single beam can be represented as a circular sector whose radius and central angle are equal to the distance and the HPBW. The arc of the sector denotes the beam coverage, where it is 2.8 m in~\cite{Orikumhi2019LocationawareBA}. Assuming the usage of the global positioning system (GPS), the root mean square (RMS) location error can be given as 15 m~\cite{736347}. Since the beam coverage is relatively small compared to the location error, the algorithm yields a large codebook despite using the location.

In this paper, the experiments are conducted indoors since the transmit power is not sufficient enough for the outdoor experiment. ULAs on both BS and UE produce a beam whose HPBW is 
$8^{\circ}$~\cite{Datasheet_ULA_Antenna}, and the distance between them is 6 m.
In this case, the beam coverage is 0.84 m.
Since the RMS location error induced by indoor positioning module is maximum 0.08 m, the beam coverage is much larger than the location error unlike the outdoor scenario. Due to this reason, we narrow down the codebook as in Section~\ref{2}, and the influence of the location error is not heavily considered.

In spite of aforementioned limitation, the experiment in this paper is meaningful since it is the first experiment that employ the location information to beam alignment. By showing the significant decrease in time consumption, this research suggests that employing the external information, such as location can be strong option for the fast beam alignment. In future, we will extend this work to outdoor and deal with unreliable location information.

\section{Conclusion}
We have experimented the location-aware beam alignment on real testbed. 
We construct MIMO testbed which generates the 13.8 GHz signal and steers the beam with the ULAs on transmitter and receiver. 
The software determines the reduced codebook based on the location information and selects the optimal beam by partial searching.   
Experiment results present that the location-aware beam tracking successfully reduce the time consumption for making beam aligned. 

\newpage

\bibliographystyle{IEEEbib}
\bibliography{reference}

\end{document}